\documentclass[aip]{revtex4}

\usepackage[usenames,dvipsnames]{color}
\usepackage{amsfonts,amssymb,amsmath}
\usepackage{bm}
\usepackage[a4paper]{geometry}
\usepackage{pdflscape}

\newcommand{\ket}[1]{\left | #1 \right \rangle}
\newcommand{\bra}[1]{\left \langle #1 \right |}
\newcommand{\braket}[2]{\langle#1|#2\rangle}
\newcommand{\proj}[1]{\ket{#1}\bra{#1}}

\definecolor{darkred}{rgb}{.8,0,0}

\definecolor{darkblue}{rgb}{0,0,.7}

\newcommand{\vz}[1]{{{#1}}}

\begin{document}

\title{Path-integral approach to the Wigner--Kirkwood expansion}

\author{Petr  Jizba$^{1,}$}
 \email{p.jizba@fjfi.cvut.cz}
\author{V\'{a}clav Zatloukal$^{2,}$}%
 \email{v.zatloukal@fu-berlin.de}
\hspace{3mm}
\affiliation{\hspace{1mm}\\ $^{1,2}$FNSPE, Czech Technical University in Prague,
B\v{r}ehov\'{a} 7, 115 19 Praha 1, Czech Republic\\  \\
$^1$ITP, Freie Universit\"{a}t in Berlin, Arnimallee 14, D-14195 Berlin, Germany \\  \\
$^2$Max Planck Institute for the History of Science, Boltzmannstrasse 22, 14195 Berlin, Germany}%



\begin{abstract}
We study the high-temperature behavior of quantum-mechanical path integrals. Starting from the Feynman--Kac
formula, we derive a new functional representation of the Wigner--Kirkwood perturbation expansion for
quantum Boltzmann densities.
As shown by its applications to different potentials, the presented expansion
turns out to be quite efficient in generating analytic form of the higher-order expansion coefficients. To put some flesh on
the bare bones we apply the expansion to obtain basic thermodynamic functions of
the one-dimensional anharmonic oscillator. Further salient issues, such as generalization to the Bloch density matrix and
comparison with the more customary world-line formulation are discussed.
The paper is accompanied by Mathematica code that generates higher order expansion terms for an arbitrary smooth local potential.
\end{abstract}

\maketitle


\section{Introduction}

The Wigner--Kirkwood (WK) expansion was originally presented in two seminal papers~\cite{W-K,W-K2} and since its very
inception it has had two important implications.
On the one hand, it has been used for studying the equilibrium statistical mechanics of a nearly classical system of
particles obeying Maxwell--Boltzmann statistics.
WK expansion is in its essence an expansion of the quantum Boltzmann density
in powers of Planck's constant $\hbar$, or equivalently of the thermal de Broglie wavelength $\lambda = \hbar\sqrt{\beta/M}$, where $\beta$
is the inverse temperature and $M$ is the
mass of a particle. On the other hand, it has paved a way for new alternative mathematical techniques and practical calculational
schemes that are
pertinent to the high-temperature regime in quantum systems.

In this paper, we pursue the study of the WK perturbation method by means of the path integral (PI) calculus.
The relevance of the PI treatment
in a high-temperature context is due to several reasons: PI's allow to connect evolutionary equations (Bloch
equation or Fokker--Planck equation) with the underlying stochastic analysis~\cite{Simon:79,Haba:99}, they are
tailor-made for obtaining quasi-classical asymptotics~\cite{Kleinert,zinn-justin}, they allow to utilize some
powerful transformation techniques to simplify the original stochastic process~\cite{fischer:92,Kleinert}, etc. Besides, PI's
also provide an excellent tool for direct numerical simulations of the underlying stochastic dynamics including
many-body systems~\cite{Ceperley:2013,Krauth:2006}. One of the key advantages of the PI approach is, however,
the fact that the techniques and methodologies used can efficiently bypass the explicit knowledge of the exact
energy spectrum --- the point that hindered earlier attempts to go beyond few leading orders in the expansion
(see, e.g., Refs.~\cite{Hioe:75,Hioe:78,Banerjee:78,Witschel:80,Schwarz}).
In particular, one can progress without relying on the explicit use of approximate expressions or interpolation
formulas for the energy eigenvalues which are often difficult to judge due to lack of reliability in their error
estimates.

The idea to use PI's as a means of producing various WK-type expansions and related thermodynamic functions is
clearly not new. Indeed, the first systematic discussions and analyzes of these issues emerged already during
the early 1970's. Among these belong the early attempts of PI treatments of the high-temperature behavior of partition
functions for anharmonic oscillators~\cite{Miller:71,Jorish:75,Korsch:79} and gradient expansions of free energy~\cite{Kleinert}.
These approaches belong in the class of the so-called {\em analytic perturbation schemes} which account for an explicit analytic
expressions of the coefficient functions.
For many practical purposes it is desirable to have explicit analytical expressions for coefficients in the
WK perturbation expansion.
This is so, for instance, when the symmetry (Lorentz, gauge, global) is supposed to be broken by quantum or thermal fluctuations.
Though these issues are more pressing in quantum field theories, they have in recent two decades entered also in the realm of a few-body
finite-temperature quantum mechanics. The catalyst has been theoretical investigations and ensuing state-of-the-arts experiments in
condensed Bose gases, degenerate Fermi gases, quantum clusters or strongly coupled Coulomb systems.
It is not only the zero-temperature regime that is of interest in these systems. Many issues revolve also around finite-temperature
or ``high''-temperature questions. These include, thermal and thermoelectric transport of ultra-cold atomic gases~\cite{Brantut,Grenier},
hydrogen, helium, and hydrogen/helium mixtures and their astrophysical implications~\cite{Chabrier:06,Martin}, Lennard--Jones
$^3$He and $^4$He gases~\cite{Boyd:68}, etc.

Apart from the aforementioned group of PI methods there are also various non-analytic methods among which the most prominent are
computational methods, such as PI Monte Carlo and  molecular dynamics simulations~\cite{Cao:94,Militzer:01},
accompanied by a host of  PI reweighted techniques~\cite{Predescu:03}. Another important type of non-analytic
method are the approximative schemes, to which belong variational approaches~\cite{Kleinert,Feynman-Kleinert,Giachetti}
and ergodic approximations~\cite{Angel}. Nice summaries of both analytic and non-analytic PI approaches can be found, e.g.,
in Refs.~\cite{Kleinert,Krauth:2006}.

A serious weakness of existent analytic WK expansions and their various disguises (be they based on PI's or not),
resides in their inability to progress very far with the expansion order. This makes it difficult to address
thermodynamically relevant intermediate-temperature regions that is particularly pertinent in molecular and
condensed matter chemistry (binding energies, self-dissociation phenomena, order-disorder transitions, etc.). The best analytic
expansions are presently available within the framework of the world-line path integral method (known also as
the string inspired
method)~\cite{schubert}. In this approach the expansion coefficients are available up to order
$\mathcal{O}(\beta^{12})$, subject to the actual interaction potential (cf. Refs.~\cite{schubert,Fliegner-I,Fliegner-II}).
Other more conventional approaches, such as the recursive or non-recursive heat-kernel calculations~\cite{Belkov:93,Carso:90} or higher
derivative expansions by Feynman diagrams~\cite{Kikkawa:76,MacKenzie:84,Fraser:85}, achieve at best the order
$\mathcal{O}(\beta^{7})$.
The key problem is a rapid escalation in the complexity of higher-order terms which is difficult to handle without some type
of resummation. In the present paper we derive a new resummation formula that provides a rather
simple and systematic way of deriving the coefficient functions.
Its main advantages rely on both an analytic control of the
high-temperature behavior, and on an accurate description over a wide temperature range via numerical calculations
that can be simply carried out at the level of an undergraduate exercise.

The structure of the paper is as follows. To set the stage we recall in the next section some fundamentals of
PI formulations of the Bloch density
matrix and the ensuing partition function and Boltzmann density. With the help of the space-time transformation
that transforms the Wiener-process
sample paths to the Brownian-bridge sample paths  we obtain the PI that represents a useful alternative to the
original Feynman--Kac representation.
Consequently we arrive at a new functional representation of the Boltzmann density which is more suitable
for tackling the high-temperature regime
than  the genuine Wigner--Kirkwood formulation
(see Sections~\ref{II.A}, \ref{II.B} and \ref{anh osc}). While the
method resembles in principle the Wentzel--Brillouin--Kramers (WKB) solution for the transition amplitude, its details are quite different. In two associated
subsections we examine some salient
technical issues related to the low-order high-temperature expansion in one dimension.
To illustrate the potency of our approach we consider in Section~\ref{anh osc} the high-temperature
expansion of the one-dimensional anharmonic oscillator.
In particular, we perform the Boltzmann density and ensuing partition function expansions and compute
the related thermodynamic quantities. The expansions
obtained improve over the classic results of Schwarz~\cite{Schwarz} and Pad\'{e}-approximation-based
expansion of Gibson~\cite{Gibson:1984}.
In Section~\ref{off-diag} we proceed by extending our expansion to the whole Bloch density matrix.
The expansion thus obtained is  compared with the
more conventional Wick's theorem based perturbation expansion based on the Onofri--Zuk Green's functions.
There we show that our prescription comprises
substantially less (in fact, exponentially less) terms
contributing to higher perturbation orders. Also the algebraic complexity of the coefficient functions involved is substantially lower in our approach. Various remarks
and generalizations are proposed in the concluding section. For the reader's convenience the paper
is supplemented with two appendices which clarify some
finer technical details. The paper is also accompanied by Mathematica code that generates the
higher-order expansion terms for arbitrary smooth local
potentials up to $18$th order in $\beta$.

Let us add a final note. Most of the presented mathematical
derivations are of a heuristic
nature --- as it should be expected from the mathematical analysis based on
the path-integral calculus. For example, it is
assumed throughout that the expansions such as (\ref{expan1}) or (\ref{expan2a}) have meaning
and represent, at least asymptotically, convergent series. Further, in a number of places,
we assume that integration and summation may be
interchanged. The basic purpose of this paper is to
find explicit formulas for the coefficient functions
$Q({\bm m}_1,\ldots,{\bm m}_n)$, and in doing so to reveal the
elaborate algebraic and combinatorial structure
present in these functions. A more rigorous treatment of
the aforementioned mathematical aspects is possible, but would involve different
language and techniques than are employed in this paper.

\section{Wigner--Kirkwood expansion \label{WKa}}

In this section we derive the Wigner--Kirkwood expansion by means of path-integral techniques. To this end we consider a $D$-dimensional non-relativistic quantum mechanical system described with the Hamiltonian
\begin{equation}
\hat{H} \ = \ \sum_{j=1}^D \frac{\hat{p}_j^2}{2 M_j} + V(\hat{\bm x}) ~,
\end{equation}
where $V({\bm x})$ is a generic smooth potential, and $\hat{p}_j = -i\hbar \frac{\partial}{\partial x_j}$. We define the Gibbs operator $e^{-\beta \hat{H}}$, where $\beta=1/(k_B T)$ is the inverse temperature, and $k_B$ the Boltzmann constant.
The partition function $Z(\beta)$ is defined as the trace of the Bloch (or canonical) density matrix, i.e., in the position representation we have the formula
\begin{equation} \label{PF1}
Z(\beta) \ = \ \int_{\mathbb{R}^D} d{\bm x} \bra{\bm x} e^{-\beta \hat{H}} \ket{\bm x} \ = \  \int_{\mathbb{R}^D} d{\bm x}  \, \ \varrho({\bm x}, \beta)\, .
\end{equation}
For brevity, we use here and throughout the convention $d{\bm x} \equiv d^D x$. The un-normalized probability density  $\varrho({\bm x}, \beta)$ is also known as the Boltzmann density.

Matrix elements of the Bloch density matrix can be represented by the path integral as~\cite{FH,Kleinert}
\begin{equation} \label{PIx}
\bra{{\bm x}_b} e^{-\beta \hat{H}} \ket{{\bm x}_a} \ = \ \int_{{\bm x}(0)={\bm x}_a}^{{\bm x}(\beta\hbar)={\bm x}_b} \!\!\!\mathcal{D}{\bm x}(\tau) \exp\left\{-\frac{1}{\hbar} \int_0^{\beta\hbar} d\tau \left[\sum_{j=1}^D \frac{M_j}{2} \ \! \dot{x}_j^2(\tau) + V({\bm x}(\tau)) \right] \right\} ~.
\end{equation}
which can be viewed as the Wick-rotated quantum-mechanical transition amplitude. Indeed, if one changes the time $\tau$ in $i\tau$, one recovers the usual transition amplitude $\langle {\bm x}_b, \tau_b| {\bm x}_a, \tau_a\rangle$
satisfying the Sch\"{o}dinger equation with the Hamiltonian $\hat{H}$ (cf. e.g.  Refs.~\cite{Kleinert,zinn-justin}). In the literature on stochastic processed is the path-integral representation of the Bloch density matrix also know as the Feynman--Kac formula~\cite{simons}.

%

For the purpose of the density matrix computation, we shall primarily consider here only diagonal matrix elements, i.e., case when ${\bm x}_b = {\bm x}_a$. We shall briefly return to the off-diagonal matrix elements in Section~\ref{off-diag}.
To proceed we perform a change of space and time variables,
${\bm x} \rightarrow {\bm x}_a + \Lambda {\bm \xi}$, $\tau \rightarrow \beta\hbar s$,
where $\Lambda$ is a diagonal matrix ${\rm diag}(\lambda_1,\ldots,\lambda_D)$ with
entries $\lambda_j = \sqrt{\beta\hbar^2/M_j}$ (corresponding to the thermal de Broglie
wavelength of the $j$th degree of freedom). The ensuing path integral
\begin{equation} \label{PIxi}
\bra{{\bm x}_a} e^{-\beta \hat{H}} \ket{{\bm x}_a} \ = \ \frac{1}{{\rm det} \Lambda} \int_{{\bm \xi}(0)={\bm 0}}^{{\bm \xi}(1)={\bm 0}} \mathcal{D}{\bm \xi}(s) \exp\left\{-\int_0^{1} ds \left[\frac{1}{2}\ \! \dot{\bm \xi}^2(s) + \beta V({\bm x}_a + \Lambda{\bm \xi}(s)) \right] \right\}\, ,
\end{equation}
is formulated in terms of dimesionless time $s$ and position ${\bm \xi}$. Note that  the size of quantum fluctuations is
now controlled by the parameters $\lambda_j$, i.e., the only place (apart from the overall PI normalization factor) where
the measure of quantum fluctuations --- $\hbar$ is present. Since $\beta$ and $\hbar^2$ appear in (\ref{PIxi}) on the same
footing, the small regime allows to treat in an unified manner both the semiclassical (small $\hbar$) and/or high-temperature
(small $\beta$) approximations.  By assuming small $\Lambda$  the potential term can be Taylor-expanded as
\begin{equation}
V({\bm x}_a + \Lambda{\bm \xi}(s)) \ = \ V({\bm x}_a) + \sum_{{\bm m} \neq {\bm 0}} \frac{V^{({\bm m})}({\bm x}_a)}{{\bm m}!} \ \!(\Lambda {\bm \xi}(s))^{\bm m} ~,
\end{equation}
where the $D$-dimensional index ${\bm m}=(m^1,\ldots,m^D)$ runs through all choices of $m^j$'s $\in \{0,\ldots,\infty\}$ except for $(m^1,\ldots,m^D) = (0,\ldots,0)$. The multi-derivative $({\bm m})$ is defined through the identity
\begin{eqnarray}
V^{({\bm m})}({\bm x}_a) \ = \ \left. \frac{\partial^{|{\bm m}|} V({\bm x})}{\partial {\bm x}^{\bm m} } \right|_{{\bm x}={\bm x}_a} \ \equiv \ \left. \frac{\partial^{m^1+\ldots+m^D} V({\bm x})}{\partial x_1^{m^1} \ldots \partial x_D^{m^D}} \right|_{{\bm x}={\bm x}_a}\,
,
\end{eqnarray}
with $|{\bm m}| = m^1+\ldots+m^D$. Finally, the multi-factorial ${\bm m}! \equiv m^1! \ldots m^D!$, and the multi-power of a $D$-dimensional vector ${\bm v}$ is defined componentwise as ${\bm v}^{\bm m} \equiv v_1^{m^1} \ldots v_D^{m^D}$.
Expanding the exponential, followed by some rearrangement, allows to cast (\ref{PIxi}) in the form
\begin{equation} \label{expan1}
\bra{{\bm x}_a} e^{-\beta \hat{H}} \ket{{\bm x}_a} =
\frac{e^{-\beta V({\bm x}_a)}}{{\rm det} \Lambda} \sum_{n=0}^\infty (-\beta)^n\!\!\!\!\!\!
\sum_{{\bm m}_1,\ldots,{\bm m}_n \neq {\bm 0}}
\prod_{j=1}^D \lambda_j^{m_1^j+\ldots+m_n^j}
\frac{V^{({\bm m}_1)}({\bm x}_a) \ldots V^{({\bm m}_n)}({\bm x}_a)}{{\bm m}_1! \ldots {\bm m}_n!}
\ \!\bar{Q} ~.
\end{equation}
At this point, we have introduced the dimensionless quantity
\begin{eqnarray} \label{Q0}
\mbox{\hspace{-0.0cm}}\bar{Q}({\bm m}_1,\ldots,{\bm m}_n) &=&
\frac{1}{n!} \int_{0}^{1} ds_1 \ldots ds_n
\int_{{\bm \xi}(0)={\bm 0}}^{{\bm \xi}(1)={\bm 0}} \mathcal{D}{\bm \xi}(s)
{\bm \xi}^{{\bm m}_1}(s_1) \ldots {\bm \xi}^{{\bm m}_n}(s_n)
\exp\left[-\int_0^{1}\!\! ds \ \! \frac{1}{2} \ \! \dot{\bm \xi}^2(s) \right] ,\nonumber \\[1mm]
&= & \frac{1}{n!} \int_{0}^{1} ds_1 \ldots ds_n \ \! \langle {\bm \xi}^{{\bm m}_1}(s_1) \ldots {\bm \xi}^{{\bm m}_n}(s_n) \rangle\, ,
\end{eqnarray}
that does not depend on physical constants or parameters of the system. $\bar{Q}$ is also manifestly symmetric under any permutation of its arguments. Let us stress that the $n=0$ term in the expansion (\ref{expan1}) equals $1$.

In (\ref{Q0}) we have denoted with $\langle \cdots \rangle$ the $(|{\bm m}_1|+\ldots+|{\bm m}_n|)$-point correlation function.
It can be evaluated using diagrammatic approach based on the so-called  Onofri--Zuk (or ``world-line'') Green's function~\cite{Onofri:78,Zuk:85,schubert,zinn-justin}
\begin{equation}
\Delta_{i j}(t,u) = -\frac{1}{2} \delta_{i j} \left[ |t-u| - (t+u-2 t u) \right] .
\label{II.9.a}
\end{equation}
We shall briefly sketch this approach in Sec.~\ref{off-diag} in connection with the off-diagonal density matrix elements.
At any rate, procedure based on Green's function (\ref{II.9.a}) proves rather impractical when higher-order terms are to be calculated. Inasmuch as we shall follow a different route. To this end we rewrite expression (\ref{Q0}) as a sum of $n!$ integrals over time-ordered sets $s_1 < \ldots < s_n$, slice the path integral at corresponding time instances, and replace the free-particle path integrals by more compact braket notation by virtue of (\ref{PIx}). We obtain
\begin{equation} \label{Q01}
\bar{Q}({\bm m}_1,\ldots,{\bm m}_n) \ = \
\frac{1}{n!} \sum_{\sigma \in S_n} Q({\bm m}_{\sigma(1)},\ldots,{\bm m}_{\sigma(n)})\,  ,
\end{equation}
where the sum runs over all permutations of $n$ indices, and
\begin{equation} \label{Q1}
Q({\bm m}_1,\ldots,{\bm m}_n) \ = \ \int_{0 < s_1 < \ldots < s_n < 1} \!\!\!\!\!\!\!\!\!\!\!\!\!\!\!\!\!\!\!\!\!\!\!\!\!\!ds_1 \ldots ds_n
\int_{\mathbb{R}^D} d{\bm y}_1 \ldots d{\bm y}_n
\prod_{\nu=0}^{n} \bra{{\bm y}_{\nu+1}} \exp\left[- (s_{\nu+1} - s_{\nu}) \frac{\hat{\bm q}^2}{2} \right] \ket{{\bm y}_\nu}
{\bm y}_\nu^{{\bm m}_\nu} ~.
\end{equation}
In the preceding we have defined ${\bm m}_0={\bm 0}$, $s_0 = 0$, $s_{n+1} = 1$, ${\bm y}_0 = {\bm y}_{n+1} = {\bm 0}$, and the momentum $\hat{\bm q} = (\hat{q}_1,\ldots,\hat{q}_D)$, conjugated to the (dimensionless) position operator $\hat{\bm y} = (\hat{y}_1,\ldots,\hat{y}_D)$. Here and throughout we use the standard convention: $\hat{q}_j = -i\frac{\partial}{\partial y_j}$ and $\braket{\bm y}{\bm q} = {e^{i {\bm q} {\bm y}}}/{(2\pi)^{D/2}}$.

Combinatorial complexity can be reduced significantly by observing that for any function $F({\bm m}_1,\ldots,{\bm m}_n)$ the following identity holds:
\begin{equation}
\sum_{{\bm m}_1,\ldots,{\bm m}_n \neq {\bm 0}}
\frac{1}{n!} \sum_{\sigma \in S_n} F({\bm m}_{\sigma(1)},\ldots,{\bm m}_{\sigma(n)})
\ = \ \sum_{{\bm m}_1,\ldots,{\bm m}_n \neq {\bm 0}} F({\bm m}_1,\ldots,{\bm m}_n)\, .
\end{equation}
This statement is not trivial, since $F$ is not supposed to be invariant under permutations of the ${\bm m}$'s. When applied to (\ref{expan1}) for
\begin{equation}
F({\bm m}_{\sigma(1)},\ldots,{\bm m}_{\sigma(n)})\ = \
\prod_{j=1}^D \lambda_j^{m_1^j+\ldots+m_n^j}
\frac{V^{({\bm m}_1)}({\bm x}_a) \ldots V^{({\bm m}_n)}({\bm x}_a)}{{\bm m}_1! \ldots {\bm m}_n!}\ \!
Q({\bm m}_{\sigma(1)},\ldots,{\bm m}_{\sigma(n)})\, ,
\end{equation}
the expansion can be then reduced to
\begin{equation} \label{expan2a}
\bra{{\bm x}_a} e^{-\beta \hat{H}} \ket{{\bm x}_a} \ = \
\frac{e^{-\beta V({\bm x}_a)}}{{\rm det} \Lambda} \sum_{n=0}^\infty (-\beta)^n\!\!\!\!\!\!
\sum_{{\bm m}_1,\ldots,{\bm m}_n \neq {\bm 0}}
\prod_{j=1}^D \lambda_j^{m_1^j+\ldots+m_n^j}
\frac{V^{({\bm m}_1)}({\bm x}_a) \ldots V^{({\bm m}_n)}({\bm x}_a)}{{\bm m}_1! \ldots {\bm m}_n!}
\ \!Q \, .
\end{equation}

This result is new. In addition, in Appendix we derive a new explicit expression for the coefficients $Q$ which proves to be very useful in the
determination of the higher-order terms. In particular, there we show that
\begin{equation} \label{Q11}
Q({\bm m}_1,\ldots,{\bm m}_n) \ = \ K
\int_{\mathbb{R}^D} \frac{d{\bm q}}{(2\pi)^D}
\left( \frac{i^{|{\bm m}_n|}}{1+\frac{{\bm q}^2}{2}} \frac{\partial^{|{\bm m}_n|}}{\partial {\bm q}^{{\bm m}_n}} \right)
\ldots
\left( \frac{i^{|{\bm m}_1|}}{1+\frac{{\bm q}^2}{2}} \frac{\partial^{|{\bm m}_1|}}{\partial {\bm q}^{{\bm m}_1}} \right)
\frac{1}{1+\frac{{\bm q}^2}{2}}
~,
\end{equation}
where the multiplicative constant has the form
\begin{equation} \label{K1}
K \ = \ \frac{ 1 }{ \Gamma\left(n+1-\frac{D}{2}+\frac{|{\bm m}_1|+\ldots+|{\bm m}_n|}{2}\right)}.
\end{equation}
From Appendix~B we can observe that the integral (\ref{Q11}) suffers the infrared divergencies precisely in those instances when the $\Gamma$-function in $K$ has pole. Consequently, in practical applications one should appropriately regularize (e.g., via dimensional regularization) both $K$ and integral in (\ref{Q11}) in order to resolve the indeterminate form of the product.

In passing we may note that because $Q$ is real, it must be equal to zero
when $|{\bm m}_1|+\ldots+|{\bm m}_n|$ together with all partial sums $m_1^j+...+m_n^j$ ($j = 1, \ldots, D$) is an even number (cf. also Section~\ref{off-diag} and Appendix~B).
So the expansion of the density matrix (\ref{expan2a}) can be reorganized as an expansion in $\hbar^2$. This is
emblematic of the Wigner--Kirkwood expansion~\cite{W-K} for systems with differentiable potentials. In the case of the
non-differentiable
potentials (cavities, billiards, etc.) the generalized derivative of Schwartz must be used instead~\cite{Vladimirov}.


Result (\ref{expan2a}) might be used for calculating higher-order terms beyond $\hbar^2$-correction
(terms up to order $\hbar^6$ have been already determined in the literature~\cite{Kiha:24}). Moreover,
the structure of (\ref{expan2a})
clearly emphasizes that expansion  is appropriate only when the involved thermal de
Broglie wavelengths are much smaller than the typical length of variation of the potential.

\subsection{Calculation of low-order terms in $D$ dimensions}\label{II.A}

In order to get further insight into structure of (\ref{expan2a}) we will now calculate first few terms
in the expansion.
To this end, we notice that a typical term in (\ref{Q11}) has the generic structure
\begin{align} \label{int1}
\int_{\mathbb{R}^D} \frac{d{\bm q}}{(2\pi)^D}
 \frac{q_1^{2r_1} \dots q_D^{2r_D}}{\left( 1+\frac{{\bm q}^2}{2} \right)^s}
&\ = \ \int_0^\infty d\sigma \frac{\sigma^{s-1}}{(s-1)!} e^{-\sigma}
\int_{\mathbb{R}^D} \frac{d{\bm q}}{(2\pi)^D}
 q_1^{2r_1} \dots q_D^{2r_D}  e^{-\sigma \frac{{\bm q}^2}{2}} \nonumber\\
&\ = \ \frac{\Gamma(s-\frac{D}{2}-|{\bm r}|)}{(s-1)! (2\pi)^{D/2} 2^{|{\bm r}|}} \prod_{j=1}^D \frac{(2r_j)!}{r_j!}
\, ,
\end{align}
where $r_1,\ldots,r_D,s \in \mathbb{N}$. If the power of any $q_j$ is odd, the above integral obviously vanishes.
For the sake of simplicity, the discussion here will be restricted to the orders in $\mathcal{O}(\beta^3)$, but it can be naturally extended to higher orders (cf. next section). At this level, we need to know (\ref{Q11}) for $n =1$ and $n=2$.

\vspace{2mm}
\noindent \textbf{Case $n=1$:} Here the lowest-order non-trivial contribution comes from $|{\bm m}_1| = 2$, with $m_1^i = \delta_{i j} + \delta_{i k}$. After differentiating
\begin{equation}
\frac{\partial^2}{\partial q_j \partial q_k} \frac{1}{1+\frac{{\bm q}^2}{2}}
\ = \ -\frac{\delta_{j k}}{\left( 1+\frac{{\bm q}^2}{2} \right)^2} + \frac{2 q_j q_k}{\left( 1+\frac{{\bm q}^2}{2} \right)^3}
\, ,
\end{equation}
we can use the formulas (\ref{Q11}) and (\ref{int1}) to find
\begin{equation}
Q({\bm m}_1) \ = \ \frac{1}{(2\pi)^{D/2}} \frac{\delta_{j k}}{6} \, .
\label{A.20}
\end{equation}

\vspace{2mm}
\noindent \textbf{Case $n=2$:}  Here the lowest-order non-trivial contribution comes from $|{\bm m}_1| = |{\bm m}_2| = 1$, with $m_1^i = \delta_{i j}$, $m_2^i = \delta_{i k}$. Consequently, we need to estimate
\begin{equation}
\frac{\partial}{\partial q_k} \left( \frac{1}{1+\frac{{\bm q}^2}{2}} \frac{\partial}{\partial q_j} \frac{1}{1+\frac{{\bm q}^2}{2}} \right)
\ = \ -\frac{\delta_{j k}}{\left( 1+\frac{{\bm q}^2}{2} \right)^3} + \frac{3 q_j q_k}{\left( 1+\frac{{\bm q}^2}{2} \right)^4}\,  ,
\end{equation}
which gives
\begin{equation}
Q({\bm m}_1,{\bm m}_2) \ = \ \frac{1}{(2\pi)^{D/2}} \frac{\delta_{j k}}{24}\,  .
\label{A.21}
\end{equation}

By gathering the results (\ref{A.20}) and (\ref{A.21}) together we can write the expansion (\ref{expan2a}) in the $n=2$ approximation as
\begin{equation} \label{expan2}
\bra{{\bm x}_a} e^{-\beta \hat{H}} \ket{{\bm x}_a} \ \sim \
\frac{e^{-\beta V({\bm x}_a)}}{ (2\pi)^{D/2} {\rm det} \Lambda }
\left(
1 - \frac{\beta}{12} \sum_{j=1}^D \lambda_j^2 \frac{\partial^2 V({\bm x}_a)}{\partial x_j^2}
+ \frac{\beta^2}{24} \sum_{j,k=1}^D \lambda_j \lambda_k \frac{\partial V({\bm x}_a)}{\partial x_j} \frac{\partial V({\bm x}_a)}{\partial x_k}
\right) .
\end{equation}
This agrees, for $\lambda_j=\lambda$ (i.e., for equal-mass particles), with the usual low-order Wigner--Kirkwood expansion (see, e.g., Refs.~\cite{landau,Kleinert}).


\subsection{Expansion for $D=1$}\label{II.B}

Here we show that the form of the coefficients $Q(m_1,\ldots,m_n)$ can be substantially simplified  in $1$-dimension ($D=1$). It is rather interesting that the simplification basically involves only arithmetic operations. To see what is
involved we denote
\begin{eqnarray}
&&I(m_1,\ldots,m_n | r,s) \ = \
\int_{\mathbb{R}} \frac{dq}{2\pi}
\left( \frac{i^{{m}_n}}{1+\frac{q^2}{2}} \frac{\partial^{{m}_n}}{\partial {q}^{{m}_n}} \right)
\ldots
\left( \frac{i^{{m}_1}}{1+\frac{q^2}{2}} \frac{\partial^{{m}_1}}{\partial {q}^{{m}_1}} \right)
\left( \frac{1}{(1+\frac{i}{\sqrt 2} q)^r} \frac{1}{(1 -\frac{i}{\sqrt 2} q)^s} \right) ,\nonumber \\[1mm]
&&~
\end{eqnarray}
so that (cf. Eq.~(\ref{Q11})):  $Q(m_1,\ldots,m_n) = K(m_1,\ldots,m_n) I(m_1,\ldots,m_n | 1,1)$. By the $m_1$-fold differentiation of the last bracket we obtain the recurrence relation
\begin{align} \label{rec1}
I(m_1,\ldots,m_n | r,s) &\ = \ \frac{(-1)^{m_1}}{2^{m_1/2}} m_1! \sum_{k_1=0}^{m_1} (-1)^{k_1} {r-1+k_1 \choose r-1} {s-1+m_1-k_1 \choose s-1} \nonumber\\ &\ \times \ I(m_2,\ldots,m_n | r+1+k_1,s+1+m_1-k_1) \, ,
\end{align}
with the initial condition
\begin{equation} \label{ini}
I(\emptyset | r,s) \ = \ \int_{\mathbb{R}} \frac{dq}{2\pi} \frac{1}{(1+\frac{i}{\sqrt 2} q)^r} \frac{1}{(1 -\frac{i}{\sqrt 2} q)^s}
\ = \ \frac{2^{3/2}}{2^{r+s}} {r+s-2 \choose r-1} \, .
\end{equation}
The latter identity is a straightforward consequence of Cauchy's integral theorem where the contour integration is taken at either the pole $i\sqrt{2}$ or $-i\sqrt{2}$.
Repeated use of (\ref{rec1}), with (\ref{ini}) as the last step, leads to an explicit form for $Q(m_1,\ldots,m_n)$, namely
\begin{equation} \label{Q3}
Q \ = \ \frac{(\frac{m_1+\ldots+m_n}{2}+n)!}{\sqrt{2\pi} 2^{(m_1+\ldots+m_n)/2}}
\sum_{\ell_1=0}^{m_1} \ldots \sum_{\ell_n=0}^{m_n}
\prod_{k=1}^{n} \frac{(-1)^{\ell_k} {m_k \choose \ell_k}}{(\ell_1\!+\!\ldots\!+\!\ell_k\!+\!k)(m_1\!-\!\ell_1\!+\!\ldots\!+\!m_k\!-\!\ell_k\!+\!k)}
~.
\end{equation}
In deriving we have used the duplication formula~\cite{Gradshteyn}: $\sqrt{\pi} 2^{1-2z} \Gamma(2z) = \Gamma(z) \Gamma(z+1/2)$.
Resulting one-dimensional expansion takes the form
\begin{equation} \label{1D}
\bra{x_a} e^{-\beta \hat{H}} \ket{x_a} \ = \
\frac{e^{-\beta V(x_a)}}{\lambda} \sum_{n=0}^\infty (-\beta)^n\!\!\!\!
\sum_{m_1,\ldots,m_n = 1}^{\infty}
\lambda^{m_1+\ldots+m_n}
\frac{V^{(m_1)}(x_a) \ldots V^{(m_n)}(x_a)}{m_1! \ldots m_n!} \ \!Q \, .
\end{equation}
\vz{Apart from the initial constant term $\beta^0 \hbar^0$, the latter expansion contains terms proporional to $\beta^i (\hbar^2)^j$, where $i,j \in \mathbb{N}$ and $i/3 \leq j \leq i-1$ (or, equivalently, $j+1 \leq i \leq 3j$).}

For the first few orders the coefficients of the expansion can be found rather straightforwardly. In Table~\ref{tab:coef} we list the coefficients of the series $e^{\beta V(x)} \sqrt{2\pi}\lambda \bra{x} e^{-\beta \hat{H}} \ket{x}$. To order $\beta^8$ these can be obtained without any excessive
hardship (for further comments see~\cite{fn1}). The higher orders in fixed
$\beta$ can now be simply obtained by grouping terms with equal order of $\beta$
and performing a number of multi-differentiations for $V(x_a)$
which can be easily done with Maple or Mathematica.
%
To this end we supplement the paper with Mathematica code that allows to generate the higher-order expansion terms (up to $18$th order) for arbitrary smooth local potentials.

%
\begin{table}
\caption{Coefficients of the series $e^{\beta V(x)} \sqrt{2\pi}\lambda \bra{x} e^{-\beta \hat{H}} \ket{x}$, at terms $\beta^i (\hbar^2)^j$, calculated according to formulas (\ref{Q3}) and (\ref{1D}). \vz{Here $0 \leq i \leq 8$ and $0 \leq j \leq 7$, which allows to determine the series up to the 8th order in $\beta$. Coefficients of terms $(\hbar^2)^j$, which are polynomials in $\beta$, can be read off completely only for $j \leq 2$. (For instance, the $\hbar^6$-term is lacking a contribution from $\beta^9$.)} }
\begin{flushleft}
 \label{tab:coef}
\begin{tabular}{|c|c|c|c|c|c|c|}
\hline\hline
 & $\hbar^0$ & $\hbar^2$ & $\hbar^4$ & $\hbar^6$ & $\hbar^8$ & $\hbar^{10}$ \\ \hline
 $\beta^0$ & $1$ & $0$ & $0$ & $0$ & $0$ & $0$  \\ \hline
 $\beta^1$ & $0$ & $0$ & $0$ & $0$ & $0$ & $0$  \\ \hline
 $\beta^2$ & $0$ & $-\frac{V''(x)}{12 M}$ & $0$ & $0$ & $0$ & $0$  \\ \hline
 $\beta^3$ & $0$ & $\frac{V'(x)^2}{24 M}$ & $-\frac{V^{(4)}(x)}{240 M^2}$ & $0$ & $0$ & $0$  \\ \hline
 $\beta^4$ & $0$ & $0$ & $\begin{matrix}\frac{V''(x)^2}{160 M^2} \\ +\frac{V'(x) V^{(3)}(x)}{120 M^2} \end{matrix}$ &
   $-\frac{V^{(6)}(x)}{6720 M^3}$ & $0$ & $0$  \\ \hline
 $\beta^5$ & $0$ & $0$ & $-\frac{11 V'(x)^2 V''(x)}{1440 M^2}$ & $\begin{matrix} \frac{23 V^{(3)}(x)^2}{40320
   M^3} \\ +\frac{19 V''(x) V^{(4)}(x)}{20160 M^3} \\ +\frac{V'(x) V^{(5)}(x)}{2240 M^3} \end{matrix}$ &
   $-\frac{V^{(8)}(x)}{241920 M^4}$ & $0$  \\ \hline
 $\beta^6$ & $0$ & $0$ & $\frac{V'(x)^4}{1152 M^2}$ & $\begin{matrix} -\frac{61 V''(x)^3}{120960 M^3} \\ -\frac{43 V'(x)
   V^{(3)}(x) V''(x)}{20160 M^3} \\ -\frac{5 V'(x)^2 V^{(4)}(x)}{8064 M^3} \end{matrix}$ & $\begin{matrix} \frac{23
   V^{(4)}(x)^2}{483840 M^4} \\ +\frac{19 V^{(3)}(x) V^{(5)}(x)}{241920 M^4} \\ +\frac{11
   V''(x) V^{(6)}(x)}{241920 M^4} \\ +\frac{V'(x) V^{(7)}(x)}{60480 M^4} \end{matrix}$ &
   $-\frac{V^{(10)}(x)}{10644480 M^5}$ \\ \hline\hline
\end{tabular}
\\[3mm]
\begin{tabular}{|c|c|c|c|c|c|c|}
\hline\hline
 & $\hbar^6$ & $\hbar^8$ & $\hbar^{10}$ & $\hbar^{12}$ & $\hbar^{14}$ \\ \hline
 $\beta^7$ & $\begin{matrix} \frac{V^{(3)}(x) V'(x)^3}{2016 M^3} \\ +\frac{83 V''(x)^2 V'(x)^2}{80640 M^3} \end{matrix}$
   & $\begin{matrix}-\frac{V^{(6)}(x) V'(x)^2}{32256 M^4} \\ -\frac{V^{(3)}(x) V^{(4)}(x) V'(x)}{4480
   M^4} \\ -\frac{V''(x) V^{(5)}(x) V'(x)}{6720 M^4} \\ -\frac{31 V''(x) V^{(3)}(x)^2}{161280
   M^4} \\ -\frac{5 V''(x)^2 V^{(4)}(x)}{32256 M^4} \end{matrix}$ & $\begin{matrix} \frac{71 V^{(5)}(x)^2}{21288960
   M^5} \\ +\frac{61 V^{(4)}(x) V^{(6)}(x)}{10644480 M^5} \\ +\frac{19 V^{(3)}(x)
   V^{(7)}(x)}{5322240 M^5} \\ +\frac{17 V''(x) V^{(8)}(x)}{10644480 M^5} \\ +\frac{V'(x)
   V^{(9)}(x)}{2128896 M^5} \end{matrix}$ & $-\frac{V^{(12)}(x)}{553512960 M^6}$ & $0$ \\ \hline
 $\beta^8$ & $-\frac{17 V'(x)^4 V''(x)}{69120 M^3}$ & $\begin{matrix} \frac{1261 V''(x)^4}{29030400
   M^4} \\ +\frac{227 V'(x) V^{(3)}(x) V''(x)^2}{604800 M^4} \\ +\frac{527 V'(x)^2 V^{(4)}(x)
   V''(x)}{2419200 M^4} \\ +\frac{659 V'(x)^2 V^{(3)}(x)^2}{4838400 M^4} \\ +\frac{17 V'(x)^3
   V^{(5)}(x)}{483840 M^4} \end{matrix}$ & $\begin{matrix}-\frac{71 V^{(8)}(x) V'(x)^2}{63866880 M^5} \\ -\frac{3067
   V^{(4)}(x) V^{(5)}(x) V'(x)}{159667200 M^5} \\ -\frac{13 V^{(3)}(x) V^{(6)}(x)
   V'(x)}{950400 M^5} \\ -\frac{109 V''(x) V^{(7)}(x) V'(x)}{15966720 M^5} \\ -\frac{6353
   V''(x) V^{(4)}(x)^2}{319334400 M^5} \\ -\frac{7939 V^{(3)}(x)^2 V^{(4)}(x)}{319334400
   M^5} \\ -\frac{13 V''(x) V^{(3)}(x) V^{(5)}(x)}{394240 M^5} \\ -\frac{3001 V''(x)^2
   V^{(6)}(x)}{319334400 M^5} \end{matrix}$ & $\begin{matrix} \frac{3433 V^{(6)}(x)^2}{16605388800 M^6} \\ +\frac{1501
   V^{(5)}(x) V^{(7)}(x)}{4151347200 M^6} \\ +\frac{2003 V^{(4)}(x) V^{(8)}(x)}{8302694400
   M^6} \\ +\frac{5 V^{(3)}(x) V^{(9)}(x)}{41513472 M^6} \\ +\frac{73 V''(x)
   V^{(10)}(x)}{1660538880 M^6} \\ +\frac{V'(x) V^{(11)}(x)}{92252160 M^6} \end{matrix}$ &
   $-\frac{V^{(14)}(x)}{33210777600 M^7}$ \\ \hline\hline
\end{tabular}
\end{flushleft}
\end{table}

\section{Example: Anharmonic oscillator in $D=1$ \label{anh osc}}

In the previous section, we have seen in some detail how the coefficients
functions in the Wigner--Kirkwood expansion can be resolved in an explicit form.
The basic results there were the formulas (\ref{expan2a})--(\ref{K1}).
The expressions found are quite general, valid for any
smooth potential and in $D=1$ are analytically
accessible up to order $\beta^{18}$. Nevertheless, for consistency reasons it is useful to examine
a problem possessing an exact solution in which it is possible to find closed expressions for
the expansion coefficients. The $D=1$ harmonic oscillator provides us
with just such an exactly solvable example. Rather than starting directly with a simple harmonic oscillator,
it is instructive to start with an {\em anharmonic} oscillator first and then regain the harmonic oscillator
solution in the limit of vanishing coupling constant (i.e., zero anharmonicity limit).
In addition, the anharmonic oscillator, which can be regarded as a field theory in one dimension, has
long served as a testing ground for new ideas for solving field theories and hence is bolstered by a large
body of literature.  In this respect it is a natural model which any new approximation scheme should
address. For a definiteness we start with the anharmonic potential
\begin{equation}
V(x)\  = \  \frac{M}{2} \omega^2 x^2 + \frac{g}{4!} x^4 ,
\label{28aa}
\end{equation}
for which the high-temperature expansion (\ref{1D}) yields
\begin{eqnarray} \label{anh1}
\bra{x} e^{-\beta \hat{H}} \ket{x} &=& \frac{\exp\left[-\beta \left( \frac{M}{2}
 \omega^2 x^2 + \frac{g}{4!} x^4 \right) \right]}{\sqrt{2\pi}\lambda}
\left[
1\ - \ \frac{\beta ^2 \hbar ^2 \left(g x^2+2 M
   \omega ^2\right)}{24 M}\right.
\nonumber\\[1mm]
&+&\frac{\beta ^3
   \left(5 M x^2 \hbar ^2 \left(g x^2+6 M \omega
   ^2\right)^2-18 g \hbar ^4\right)}{4320
   M^2}\nonumber \\[1mm]
&+& \left.\frac{\beta ^4 \hbar ^4 \left(17 g^2 x^4+84
   g M x^2 \omega ^2+36 M^2 \omega ^4\right)}{5760
   M^2}\ + \ O\left(\beta ^5\right)
\right] .
\end{eqnarray}
The higher-order corrections can be explicitly obtained with the help of Table~\ref{tab:coef} (up to order $\beta^8$) or with the enclosed Mathematica code quoted in~\cite{fn1} (up to order $\beta^{18}$).

In the case of zero anharmonicity $(g=0)$, we can check our results against the exact solution of the harmonic oscillator problem. The expansion (\ref{anh1}) reduces to
\begin{eqnarray}
&&\mbox{\hspace{-5mm}}\bra{x} e^{-\beta \hat{H}} \ket{x}_{g=0}  = \ \frac{\exp\left(-\beta \frac{M}{2} \omega^2 x^2 \right)}{\sqrt{2\pi}\lambda}
\Bigg[
1-\frac{1}{12} \beta ^2 \omega ^2 \hbar
   ^2+\frac{1}{24} \beta ^3 M x^2 \omega ^4
   \hbar ^2+\frac{1}{160} \beta ^4 \omega ^4 \hbar
   ^4+O\left(\beta ^5\right)
\Bigg] ,\nonumber \\[1mm]
&&\mbox{\hspace{-5mm}}~
\end{eqnarray}
which, indeed, coincides with the corresponding expansion of the well-known analytic form of the Bloch density matrix for harmonic oscillator (see, e.g., Refs.~\cite{Kleinert,zinn-justin})
\begin{equation}
\bra{x} e^{-\beta \hat{H}} \ket{x}_{g=0} \ = \ \frac{\exp\left(-\beta \frac{M}{2} \omega^2 x^2 \right)}{\sqrt{2\pi}\lambda}
\sqrt{\frac{\beta  \omega  \hbar}{  \text{sinh}(\beta  \omega
   \hbar )}} \exp \left[-\frac{M x^2 \omega}{\hbar}  \left(\tanh
   \frac{\beta  \omega  \hbar
   }{2}-\frac{\beta  \omega  \hbar
   }{2}\right) \right]
.
\end{equation}

In passing we may note that the expansion of the single-particle partition function $Z(\beta)$ associated with
(\ref{anh1}) can be phrased in terms of the {\em parabolic cylindric function} and its derivatives which, after
re-expansion, give
%
%
\begin{eqnarray}
Z(\beta) \ &=& \  \frac{1}{\sqrt{2\pi} \lambda}  {{\sqrt[4]{\frac{3}{2 \beta g}}}} \left[
 \Gamma
   \left(\frac{1}{4}\right)  \ + \ \frac{\sqrt{\frac{3}{2}}
   \sqrt{\beta } M \omega ^2 \Gamma
   \left(-\frac{1}{4}\right)}{2 \sqrt{g}} \ + \ \frac{3
    \beta  M^2 \omega ^4
   \Gamma
   \left(\frac{5}{4}\right)}{g}\right.\nonumber \\[2mm]
   &-& \ \frac{\beta
   ^{3/2} \left(\Gamma \left(\frac{3}{4}\right)
   \left(g^2 \hbar ^2+18 M^4 \omega
   ^6\right)\right)}{4 \left(\sqrt{6}
   g^{3/2} M\right)}\nonumber \\[2mm]
   &-& \ \left. \frac{\beta ^{2}
   \left( \Gamma
   \left(-\frac{3}{4}\right) \left(2 g^2 \omega ^2
   \hbar ^2+45 M^4 \omega ^8\right)\right)}{128
   g^{2}} \ + \ \mathcal{O}\left(\beta ^{5/2}\right)
\right]\, .
\end{eqnarray}
This, when combined with appropriate thermodynamic
formulas, yields the following expressions for entropy $S$, the heat
capacity $C_V$ and internal energy $U$:
\begin{align}
\frac{S}{k_B} \ = \ - \frac{1}{k_B}\left(\frac{\partial F}{\partial T}  \right)_{V} \nonumber \ &= \  \log Z(\beta) -  \frac{\beta }{Z(\beta)} \left(\frac{\partial Z(\beta)}{\partial \beta}  \right)_V
\nonumber\\[2mm]
\ &= \ \frac{3}{4}\ + \ \log \left(\frac{2\Gamma
   \left(\frac{5}{4}\right)}{\lambda}\sqrt[4]{\frac{6}{\pi ^2
   \beta g}} \right)\ - \ \frac{\sqrt
   {\frac{3}{2}} \sqrt{\beta } M \omega ^2 \Gamma
   \left(\frac{3}{4}\right)}{\sqrt{g} \Gamma
   \left(\frac{1}{4}\right)}\nonumber \\[2mm]
   &+ \ \frac{\beta ^{3/2}
   \left(\pi  g^2 \hbar ^2 \Gamma
   \left(\frac{5}{4}\right) \ + \ 3 \sqrt{2} M^4 \omega ^6
   \Gamma \left(\frac{3}{4}\right)^3\right)}{\sqrt{3}
   g^{3/2} M \Gamma
   \left(\frac{1}{4}\right)^3} \ + \ \mathcal{O}\left(\beta ^2\right),
\nonumber\\[3mm]
\frac{C_V}{k_B} \ = \ \frac{T}{k_B} \left(\frac{\partial S}{\partial T}  \right)_V \ &= \ - \frac{\beta}{k_B} \left(\frac{\partial S}{\partial \beta}  \right)_V
\nonumber\\[2mm]
\ &= \ \frac{3}{4}\ + \ \frac{\sqrt{\frac{3}{2}} \sqrt{\beta } M
   \omega ^2 \Gamma \left(\frac{3}{4}\right)}{2
   \sqrt{g} \Gamma
   \left(\frac{1}{4}\right)}\nonumber\\[2mm]
   &- \ \frac{\beta ^{3/2}
   \left(\sqrt{3} \pi  g^2 \hbar ^2 \Gamma
   \left(\frac{5}{4}\right)+3 \sqrt{6} M^4 \omega ^6
   \Gamma \left(\frac{3}{4}\right)^3\right)}{2
   g^{3/2} M \Gamma
   \left(\frac{1}{4}\right)^3}\ + \ \mathcal{O}\left(\beta
   ^2\right),
\nonumber\\[3mm]
{U} \ = \  - {T^2} \left(\frac{\partial F/T}{\partial T} \right)_V \ &= \ \left( \frac{\partial F \beta}{\partial \beta} \right)_V
\nonumber\\[2mm]
\ &= \
 \frac{3}{4 \beta }+\frac{\sqrt{\frac{3}{2}} M \omega
   ^2 \Gamma \left(\frac{3}{4}\right)}{\sqrt{\beta }
   \sqrt{g} \Gamma \left(\frac{1}{4}\right)}-\frac{3
   M^2 \omega ^4 \left(\Gamma
   \left(\frac{1}{4}\right)^2-4 \Gamma
   \left(\frac{3}{4}\right)^2\right)}{4
   g \Gamma
   \left(\frac{1}{4}\right)^2}\nonumber\\[2mm]
&+\frac{\sqrt{\beta } \left(2 \sqrt{3} \pi  g^2 \hbar ^2 \Gamma
   \left(\frac{9}{4}\right)+15 \sqrt{6} M^4 \omega
   ^6 \Gamma \left(\frac{3}{4}\right)^3\right)}{320
   g^{3/2} M \Gamma
   \left(\frac{5}{4}\right)^3} +\ \mathcal{O}\left(\beta \right)\,  .
\end{align}
[$F = -k_B T \log Z(\beta)$ is the Helmholtz free energy]. These expansions
are not only in excellent agreement with the classic (spectral-theorem based) expansions of Schwarz~\cite{Schwarz} and Gibson~\cite{Gibson:1984} but they also
go beyond these expansions by providing explicit forms
for higher-order terms not present in Refs.~\cite{Schwarz,Gibson:1984}.

Unfortunately when $M$  in (\ref{28aa}) is  negative (i.e., we would have a double-well potential) the WK approach would fail. Indeed the WK expansion cannot accommodate non-perturbation effect such as multi-instanton contribution and ensuing tunneling, as by its very construction it is basically a perturbation expansion around a free solution. From this point of view a tunneling in a double well potential seems to be beyond reach in our expansion. Of course, tunneling could be included by considering some sort of a hybrid approach in which the ``phase part" of the transition probability would be calculated via WKB (possibly including multi-istanton contribution), while the fluctuating factor would be evaluated perturbatively via WK method. One of the potential bonuses would be the fact that one could bypass the notorious problems with the Van Vleck determinant on caustics. Such a hybrid approach would, however, clearly go beyond the simple WK approach that is used in our paper. In our future investigation we will touch more upon this issue.



%


\section{Off-diagonal Bloch density matrix elements \label{off-diag}}

So far, we have almost exclusively been dealing with the diagonal elements of the Bloch
density matrix --- Boltzmann density. This was well justified by expected applications in
statistical physics, where typically only the
partition function is required and hence only diagonal elements of the
density matrix are relevant \vz{(of course, only as long as the Maxwell-Boltzmann statistics is considered)}. This is also the linchpin  of the original
Wigner--Kirkwood work.

Expansion and the formula for the Bloch density matrix (\ref{expan2a}) can be straightforwardly generalized beyond the
original Wigner--Kirkwood analysis
by considering the off-diagonal form of the density matrix (also called the heat kernel or euclidean Feynman
amplitude). This would be particularly pertinent
in cases, when one would like to incorporate the exchange effects that are a consequence of fermion or boson
statistics or when the linear response theory would be in question. By following the same train of thought
as in Sec.~\ref{WKa} we can phrase the path-integral representation
of the full Bloch density matrix in
terms sum over the Brownian bridge sample paths. The path transformation that transforms the Wiener process
$\Omega_W = \left\{{\bm x}(\cdot)  \right\}$ to the Brownian bridge process $\Omega_{BB} = \left\{{\bm \xi}(\cdot)  \right\}$ is
\begin{eqnarray}
{\bm x}(\tau) \ = \ {\bm x}_a(1-s) + {\bm x}_b s + \Lambda{\bm \xi}(s)\, .
\end{eqnarray}
Let us recall that the ``Euclidean time" variable $\tau$ is connected with $s$ via the relation $\tau = \beta \hbar s$.

The Brownian bridge sample paths fulfill the Dirichlet boundary conditions
%
${\bm \xi}(0)  = {\bm \xi}(1)  =  {\bm 0}.$
%
With this the Feynman--Kac formula for the Bloch density matrix (\ref{PIx}) acquires the form
\begin{eqnarray}\label{V.1.a}
&&\mbox{\hspace{-16mm}}\bra{{\bm x}_b} e^{-\beta \hat{H}} \ket{{\bm x}_a} \ = \ \frac{\exp\left\{-\mbox{$\frac{1}{2}$}[\Lambda^{-1}({\bm x}_b - {\bm x}_a)]^2  \right\}}{{\rm det} \Lambda} \nonumber \\[2mm]
&&\mbox{\hspace{-5mm}}\times \ \int_{{\bm \xi}(0)={\bm 0}}^{{\bm \xi}(1)={\bm 0}} \mathcal{D}{\bm \xi}(s) \exp\left\{-\int_0^{1} ds \left[\frac{1}{2}\ \! \dot{\bm \xi}^2(s) + \beta V({\bm x}_a(1-s) + {\bm x}_b s + \Lambda{\bm \xi}(s)) \right]
\right\}\, ,
\end{eqnarray}
where the surface term in the action got canceled due to boundary conditions of the Brownian bridge.
We can expand the potential $V(\ldots)$ around the free-particle classical solution as
\begin{equation}
V({\bm x}_a(1-s) + {\bm x}_b s + \Lambda{\bm \xi}(s)) \ = \ V({\bm x}_a(1-s) + {\bm x}_b s) + \sum_{{\bm m} \neq {\bm 0}} \frac{V^{({\bm m})}({\bm x}_a(1-s) + {\bm x}_b s)}{{\bm m}!} \ \!(\Lambda {\bm \xi}(s))^{\bm m} ~,
\end{equation}
and write the density matrix in the form
\begin{eqnarray}
&&\mbox{\hspace{-10mm}}\bra{{\bm x}_b} e^{-\beta \hat{H}} \ket{{\bm x}_a}  =  \frac{\exp\left\{-\mbox{$\frac{1}{2}$}[\Lambda^{-1}({\bm x}_b - {\bm x}_a)]^2  - \beta \tilde{V}({\bm x}_b,{\bm x}_a)   \right\}}{{\rm det} \Lambda}
\nonumber \\[2mm]
&&\mbox{\hspace{-10mm}}\times \ \sum_{n=0}^\infty (-\beta)^n\!\!\!\!\!\!
\sum_{{\bm m}_1,\ldots,{\bm m}_n \neq {\bm 0}}
\prod_{j=1}^D \lambda_j^{m_1^j+\ldots+m_n^j}\ \!
\frac{(\tilde{V}^{({\bm m}_1)}_{s_1}({\bm x}_b,{\bm x}_a) \ldots \tilde{V}^{({\bm m}_n)}_{s_n}({\bm x}_b,{\bm x}_a) \ast \bar{Q}_{s_1 \cdots s_n})({\bm t} =1)}{{\bm m}_1! \ldots {\bm m}_n!}
 ~.
\label{V.35.a}
\end{eqnarray}
In the previous we have introduced the abbreviations
\begin{eqnarray}
&&\tilde{V}({\bm x}_b,{\bm x}_a) \ = \ \int_{0}^1 ds \ \! V({\bm x}_a(1-s) + {\bm x}_b s)\, ,\nonumber \\[2mm]
&&\tilde{V}^{({\bm m}_k)}_{s_i}({\bm x}_b,{\bm x}_a) \ = \ {V}^{({\bm m}_k)}({\bm x}_b  - s_i ({\bm x}_b  - {\bm x}_a ))\, .
\label{38.bc}
\end{eqnarray}
The multi-dimensional convolution appearing in (\ref{V.35.a}) is a straightforward extension of the one-dimensional
convolution
\begin{eqnarray}
(X(s_i)\ast Y(s_i))(t) \ = \ (Y(s_i)\ast X(s_i))(t)  \ = \ \int_{0}^t ds_i \ \! X(t-s_i) Y(s_i)\, .
\end{eqnarray}
It is the above definition of the convolution which dictates the (seemingly strangely appearing) form of the right-hand-side of (\ref{38.bc}).
Sub-indices $s_i$ appearing in $\bar{Q}$ in (\ref{V.35.a}) just indicate the integration variables in the convolution.
We see again, that the key  object is the coefficient function $\bar{Q}$ (cf. Eq.~(\ref{Q0})) or better the ensuing  $(|{\bm m}_1|+\ldots+|{\bm m}_n|)$-point correlator $\bar{Q}_{s_1 \cdots s_n}$
\begin{eqnarray}
&&\int_{{\bm \xi}(0)={\bm 0}}^{{\bm \xi}(1)={\bm 0}} \mathcal{D}{\bm \xi}(s)\ \!
{\bm \xi}^{{\bm m}_1}(s_1) \ldots {\bm \xi}^{{\bm m}_n}(s_n)
\exp\left[-\int_0^{1}\!\! ds \ \! \frac{1}{2} \ \! \dot{\bm \xi}^2(s) \right] \nonumber \\[1mm]
&&= \ \left. \frac{\delta^{|{\bm m}_1|+\ldots+|{\bm m}_n|}}{\delta{\bm j}(s_1)^{{\bm m}_1} \ldots \delta{\bm j}(s_n)^{{\bm m}_n}}
\int_{{\bm \xi}(0)={\bm 0}}^{{\bm \xi}(1)={\bm 0}} \mathcal{D}{\bm \xi}(s)\ \! \exp\left[-\int_0^{1}\!\! ds \ \! \frac{1}{2} \ \! \dot{\bm \xi}^2(s) + \int_0^{1}\!\! ds \ \! {\bm j}(s)\cdot {\bm \xi}(s)  \right]\right|_{{\bm j} = 0}
\nonumber \\[2mm]
&&= \ \mathcal{N} \left. \frac{\delta^{|{\bm m}_1|+\ldots+|{\bm m}_n|}}{\delta{\bm j}(s_1)^{{\bm m}_1} \ldots \delta{\bm j}(s_n)^{{\bm m}_n}} \ \! \exp\left[\frac{1}{2} \int_{0}^1 ds du \ \! j_i(s) \Delta_{ij}(s,u) j_j(u)  \right]\right|_{{\bm j} = 0}\, .
\end{eqnarray}
%
The normalization constant $\mathcal{N}$ denotes the path integral for a simple Brownian bridge. The summation convention is automatically utilized in the argument of the exponent on the last line. The Green function $\Delta_{ij}(s,u)$ is chosen so that it satisfies the equations
\begin{eqnarray}
&&\frac{\partial^2}{\partial t^2} \Delta_{ij}(t,u) \ = \ -\delta_{ij}
\delta(t-u)\, ,\nonumber \\[2mm]
&& \Delta_{ij} (0,u) \ = \ \Delta_{ij}(1,u) \ = \ 0\, .
\end{eqnarray}
The solution is the world-line Green function of Onofri and Zuk~\cite{Onofri:78,Zuk:85}
\begin{equation}
\Delta_{i j}(s,u)\ = \ -\frac{1}{2} \delta_{i j} \left[ |t-u| - (t+u-2 t u) \right] .
\label{w-l.a}
\end{equation}
As a result, we can write $\bar{Q}({\bm m}_1,\ldots,{\bm m}_n)$ in the form
\begin{eqnarray}
\mbox{\hspace{-5mm}}\bar{Q} &=&  \frac{\mathcal{N}}{n!}\int_{0}^1 ds_1 \ldots ds_n \ \! \frac{\delta^{|{\bm m}_1|+\ldots+|{\bm m}_n|}}{\delta{\bm j}(s_1)^{{\bm m}_1} \ldots \delta{\bm j}(s_n)^{{\bm m}_n}}\left. \exp\left[\frac{1}{2} \int_{0}^1 \! dt du \ \! j_i(t) \Delta_{ij}(t,u) j_j(u)  \right]\right|_{{\bm j} = 0} .
\end{eqnarray}
The former can be further simplified with the help of Coleman's identity:
\begin{eqnarray}
F(-i \partial/\partial{\bm x} ) G({\bm x}) \ = \ \left.G(-i\partial/\partial {\bm y}) F({\bm y}) e^{i {\bm y}\cdot {\bm x}}\right|_{{\bm y} = 0}\, ,
\end{eqnarray}
which is valid for any (sufficiently smooth) functions $F$ and $G$. After some additional algebra one verifies that
\begin{eqnarray}
\mbox{\hspace{-0mm}}\bar{Q}\ &=& \ \frac{\mathcal{N}\ \! i^{|{\bm m}_1|+\ldots+|{\bm m}_n|} }{n!} \ \! \exp\left[-\frac{1}{2} \int_0^1 dt du \ \! \frac{\delta}{\delta z_i(t)} \ \! \Delta_{ij}(t,u) \frac{\delta}{\delta z_j(u)}   \right]\nonumber \\[1mm] &&\times \left. \int_{0}^1 ds_1 \ldots ds_n \ \! {\bm z}(s_1)^{{\bm m}_1} \ldots {\bm z}(s_n)^{{\bm m}_n}\right|_{{\bm z} = 0}.
\label{IV.38.a}
\end{eqnarray}
For similar reasons as in ordinary quantum field theory, i.e., namely for the Wick theorem application, it might be convenient to formulate the $\bar{Q}$-function in the Fourier picture.
In this case the Fourier transform is discrete due to the Dirichlet boundary conditions for ${\bm \xi}$.
In addition, when we periodically extend the shape of the potential $V$ from the interval $s \in [0,1]$ to the whole $\mathbb{R}$
and take the Fourier transform  the calculations of (\ref{V.35.a}) will substantially simplify due to the convolution theorem~\cite{Vladimirov}.

Form (\ref{IV.38.a}) indicates that $Q$ can be calculated via Wick's theorem with world-line Green's functions
(\ref{w-l.a}), cf. also Refs~\cite{Kleinert,Bern-Kosower,schubert,Strassler}. In fact, it is
not difficult to list the corresponding Feynman-like diagrammatic rules for $\bar{Q}({\bm m}_1,\ldots,{\bm m}_n)$.
On the other hand, the number of terms involved in evaluating  $\bar{Q}$ via (\ref{IV.38.a}) grows
as $(2\mathfrak{m}-1)!! = (2\mathfrak{m})!/2^\mathfrak{m} \mathfrak{m}!$ where
$\mathfrak{m} = (|{\bm m}_1|+\ldots+|{\bm m}_n|)$ (see, for instance, Ref.~\cite{BJV}).
This should be contrasted with (\ref{Q11}) where the number of terms grows as
(see Appendix~\ref{App.2})
\begin{equation}
\prod_{j=1}^D \left[ (m_1^j+\ldots+m_n^j)/2+1 \right] .
\label{B.8b}
\end{equation}
%
Our prescription comprises substantially less terms and this is even more pronounced at high values
of ${\bm m}_j$'s (i.e., at high derivative orders). In Appendix~\ref{App.2} we prove that
the inequality
\begin{eqnarray}
(2\mathfrak{m} - 1)!! \ \geq \ \vz{\prod_{j=1}^D \left[ (m_1^j+\ldots+m_n^j)/2+1 \right]} \, ,\label{47.a}
\end{eqnarray}
always holds whenever $\mathfrak{m} \geq 2$. There we also show that the number of terms is in our case exponentially lower than in Wick's theorem based approaches.


Note, also that the number of $s$-integrations in (\ref{IV.38.a}) matches the perturbation order, i.e., $n$, while the number of integrations in our formula (\ref{Q11}) equals to the dimension of particles configuration space. In this respect, is the presented method less complex with the increasing perturbation order than other methods in use.  As a matter of fact, with the method based on the world-line Green function, a complete calculation of all coefficients was achieved to order $\mathcal{O}(\beta^{12})$, see Ref.~\cite{Schubert.3}.
Closely related gradient expansion calculations (with the same order of precision) were performed in Ref.~\cite{Schubert.2}

Finally, note that $\bar{Q}$ from ({\ref{IV.38.a}) is non-zero only when $|{\bm m}_1|+\ldots+|{\bm m}_n|$ together with all partial sums $m_1^j+...+m_n^j$ ($j = 1, \ldots, D$) is even.
In fact, also all partial sums $m_1^j+...+m_n^j$ ($j = 1, \ldots, D$) must be zero.  This
fact was already pointed out in Sect.~\ref{WKa} in connection with the coefficient $Q$. Again, the evenness is true only for smooth potentials.
In the general case the space derivatives must be substituted with the generalized derivative of Schwartz~\cite{Vladimirov} which
can bring about also odd terms\vz{, i.e. odd powers of $\hbar$. Also other non-analytical behaviour can emerge --- e.g., it was shown in~\cite{Jan78} that exchange contributions to the free energy of the jellium vanish exponentially fast with  $\hbar$ as a consequence of the Coulomb repulsion between identical charges which diverges at zero separation.}

\section{Conclusions and perspectives}

In this paper we have presented a novel PI-based high-temperature expansions of the Boltzmann's
density $\varrho({\bm x}, \beta)$ and partition function $Z(\beta)$. Ensuing
generalizations to the full Bloch density matrix were also discussed and
explicitly compared with the Onofri--Zuk world-line approach. It was found that
our prescription comprises substantially less terms contributing to
higher perturbation orders than the more conventional Wick's theorem based perturbation expansions.
Also the algebraic complexity of the coefficient functions involved is markedly lower in our approach.

The expansions obtained are valid for arbitrary
number of particles and provide an analytic control of the high-temperature behavior.
In addition, the implementation is sufficiently general for any system described by smooth
potential energy functions. Because of its analytic form, the presented high-temperature
expansion can be further conveniently used, e.g., to analyze the breakdown of symmetry,
generate a gradient expansion for the free energy for a wide class of potentials,
calculate ground-state energies, set up the extended Thomas--Fermi approximations or serve as the starting
point for a numerical evaluation of various thermodynamical quantities
(e.g., virial coefficients, specific heat or entropy). As
a demonstration, we have briefly discussed the high-temperature thermodynamics of the
anharmonic oscillator.


We would like to remark that the compactness
of our form for the coefficient function $Q$ might be deceptive
with regard to applications requiring the use of non-local potentials.
In the genuine Wigner--Kirkwood
method, the single-particle density is expressed
as functional of the one-body potential $V({\bm x})$.
Though our treatment can accommodate also few-body potentials it is intrinsically
formulated only for local potentials. Since
non-local potentials are an integral part of statistical quantum
theory, e.g. in cases when the exchange part of the Hartree--Fock
self-consistent  potential  is considered, the corresponding generalization
of Eqs.~(\ref{Q11}) and (\ref{V.35.a}) to non-local potentials
would be desirable. Situation is simple only for two-particle systems with potentials
of the form $V({\bm x}_1, {\bm x}_2) = V({\bm x}_1 - {\bm x}_2)$. There the transformation to the center of
mass frame allows to reduce the problem to a single-particle in an external potential $V({\bm x})$.
For other cases, the formula (\ref{Q11}) for the coefficient functions $Q$ could be derived in the same
spirit as in Section~\ref{WKa} but the appealing simplicity of $Q$ would be clearly lost.

The versatility of the method developed in this paper together with a renewed interest in the study of the high-temperature
asymptotic expansions of the Bloch density matrix suggest several extensions of this work. A pertinent extension could address
spin-dependent potentials, like the spin-orbit interaction whose interest
in nuclear physics is well known. Also the case of momentum dependent terms which are relevant in charged particle systems interacting with
electromagnetic field or in Brueckner's theory used in nuclear physics, would be desirable to include.
Important limitation of our method lies in the fact that our discussion was confined  only to cases where
Hamiltonians did not include fermionic degrees of freedom. Similarly as the original WK method also our approach is
inherently formulated within the framework of Boltzmann statistics  and so it does not incorporate the exchange effects
(which are relevant, e.g., in a hot Fermionic plasma). There exist various generalizations of the
WK formalism to include \vz{the effects of magnetic field~\cite{Ala79}}, or exchange corrections (see, e.g., Refs.~\cite{Wang:87,Caro:96}) and the corresponding
extension of our approach  in this direction would be also worth of pursuing particularly in view a naturalness
with which PI's handle fermionic particle systems~\cite{Kleinert}.
All these aforementioned issues are currently under active investigation.

It appears worthwhile to stressed that the WK expansion is not the WKB expansion. For instance, the leading asymptotic behaviors are different; while the WK expansion starts with $\exp(-\beta V(x))$, the WKB starts with $\exp(-\beta S[x_{\rm cl},x])$ (here  the action functional $S$ is evaluated along the classical solutions $x_{\rm cl}$ with the boundary conditions $x(0) = x(\hbar \beta) = x$). Even starting points for both these expansions were historically different. WK started with the Wigner transform approach to statistical physics~\cite{W-K,W-K2} while WKB (in PI's)
started with the expansion (in terms of moments of Gaussian fluctuations) around classical trajectory~\cite{Kleinert,zinn-justin}. It is also clear that in the WK one does not organize the expansion in terms of orders of fluctuations around classical solution (as the WKB does). Naturally, both approaches share many common features and there is a bulk of the literature comparing both methods and their respective pros and cons. The interested reader can see, e.g., Refs.~\cite{Osborn,Durand}.

Let us finally make a few comments concerning the low-temperature regime. It is clear that when the temperature decreases, the de Broglie wavelength increases, and the Wigner--Kirkwood perturbation expansion becomes unwarranted. This happens whenever the involved thermal de Broglie wavelength is comparable with a typical length over which the potential varies. So the low temperature expansion is normally beyond reach of the WK method. Nevertheless, with the high-temperature expansion at one's disposal one can tackle  also the low-temperature expansion (at least numerically) provided the sufficient number of the coefficient functions in the high-temperature series is available.
To this end, one is free to employ some of the existent duality approaches.
Among these, a particularly powerful is a nonperturbative approximation scheme called {\em variational} or {\em optimized perturbation theory}~\cite{Kleinert,kleinert_b,Bender,Stevenson}. There the basic idea is to combine the
renormalization-group concept known as the {\em principle of minimal sensitivity}~\cite{Stevenson} with
the techniques of perturbation theory and the variational principle to convert the divergent weak-coupling
power series into a convergent strong-coupling power series (and vice versa).


%
%
%


%

Last but not least,  recently Paulin {\em et al.}~\cite{Angel} employed the concept of the occupation time for Wiener
processes to formulate the so-called {\em ergodic local-time approximation} to PI's.
The ergodic approximation is particularly well suited for the low-temperature regime.
In high-temperature domain it performs less satisfactory since the non-trivial correlations between occupation
times must be taken into account~\cite{Angel}. Finding the dictionary that would allow a simple passage between
our and Paulin {\em et al.} approach in the high and intermediate-temperature regimes
would be particularly desirable in light of a similar mathematical
structure (namely Eq.~(\ref{PIxi}) that both approaches share.
Work along these lines is presently in progress.

\section*{Acknowledgments}

We would like to acknowledge helpful feedbacks
from J.~Klauder, H.~Kleinert and C.~Schubert. This work was supported
by GA\v{C}R Grant No. P402/12/J077.

\appendix
\section{Simplification of coefficients $Q({\bm m}_1,\ldots,{\bm m}_n)$}\label{App.1}

Here we employ a convenient trick that will allow us to can carry out the $s$-integrations in (\ref{Q1}) explicitly. We first formally promote the upper limit of the $s$-integrations (i.e., $1$) to a new variable $s_{n+1}$, and Laplace-transform $Q$ with respect to $s_{n+1}$, i.e.
\begin{eqnarray}
&&\widetilde{Q}(E)\ = \ \int_{0}^{\infty} ds_{n+1} e^{- E s_{n+1}}
\int_{0 < s_1 < \ldots < s_n < s_{n+1}} \!\!\!\!\!\!\!\!\!\!\!\!\!\!\!\!\!\!\!\!\!\!\!\!\!\!ds_1 \ldots ds_{n}
\int_{\mathbb{R}^D} d{\bm y}_1 \ldots d{\bm y}_n\nonumber \\[2mm]
&&\mbox{\hspace{1cm}}\times \  \prod_{\nu=0}^{n} \bra{{\bm y}_{\nu+1}} \exp\left[- (s_{\nu+1} - s_{\nu}) \frac{\hat{\bm q}^2}{2}\right] \ket{{\bm y}_\nu}
{\bm y}_\nu^{{\bm m}_\nu} \, .
\end{eqnarray}
Change of variables $s'_\nu = s_{\nu+1} - s_\nu$, $\nu=0,\ldots,n$, then leads to
\begin{equation}
\widetilde{Q}(E) \ = \
\int_{0}^\infty ds'_0 \ldots ds'_{n}
\int_{\mathbb{R}^D} d{\bm y}_1 \ldots d{\bm y}_n
\prod_{\nu=0}^{n} \bra{{\bm y}_{\nu+1}} \exp\left[-s'_\nu \left(E + \frac{\hat{\bm q}^2}{2}\right) \right] \ket{{\bm y}_\nu}
{\bm y}_\nu^{{\bm m}_\nu} ~.
\end{equation}
The $s$-integrations can now be done easily,
\begin{equation}
\widetilde{Q}(E)\ = \
\int_{\mathbb{R}^D} d{\bm y}_1 \ldots d{\bm y}_n
\prod_{\nu=0}^{n} \bra{{\bm y}_{\nu+1}} \frac{1}{E + \frac{\hat{\bm q}^2}{2}} \ket{{\bm y}_\nu}
{\bm y}_\nu^{{\bm m}_\nu} ~.
\label{A.3.a}
\end{equation}

In order to further simplify (\ref{A.3.a}) we perform the re-scaling ${\bm y}_\nu \rightarrow {\bm y}_\nu/{\sqrt{E}}$, and use the fact that
\begin{eqnarray}
&&\bra{\frac{{\bm y}_{\nu+1}}{\sqrt{E}}} \frac{1}{E + \frac{\hat{\bm q}^2}{2}} \ket{\frac{{\bm y}_\nu}{\sqrt{E}}}
\ = \ \int_{\mathbb{R}^D} \frac{d{\bm q}}{(2\pi)^D}
\frac{ \exp\left(i {\bm q} \frac{{\bm y}_{\nu+1} - {\bm y}_\nu}{\sqrt{E}} \right) }{ E + \frac{\hat{\bm q}^2}{2} }\nonumber \\[2mm]
&&\mbox{\hspace{3.3cm}}\overset{ {\bm q} \rightarrow \sqrt{E}{\bm q} }{=}
E^{D/2-1} \bra{{\bm y}_{\nu+1}} \frac{1}{1 + \frac{\hat{\bm q}^2}{2} } \ket{{\bm y}_\nu}
~.
\end{eqnarray}
This explicitly decouples $E$, giving rise to
\begin{equation}
\widetilde{Q}(E)\ = \
E^{D/2 - n-1 - (|{\bm m}_1|+\ldots+|{\bm m}_n|)/2 }
\int_{\mathbb{R}^D} d{\bm y}_1 \ldots d{\bm y}_n
\prod_{\nu=0}^{n} \bra{{\bm y}_{\nu+1}} \frac{1}{1 + \frac{\hat{\bm q}^2}{2} } \ket{{\bm y}_\nu}
{\bm y}_\nu^{{\bm m}_\nu} ~.
\end{equation}
Now the inverse Laplace transform can be performed and evaluated at $s_{n+1}=1$. With the help of the formula $\int_0^\infty ds \ \! s^\nu e^{-s E} = \Gamma(\nu+1) E^{-\nu-1}$, we obtain
\begin{equation}
Q \ = \ K
\int_{\mathbb{R}^D} d{\bm y}_1 \ldots d{\bm y}_n
\prod_{\nu=0}^{n} \bra{{\bm y}_{\nu+1}} \frac{1}{1 + \frac{\hat{\bm q}^2}{2}} \ket{{\bm y}_\nu}
{\bm y}_\nu^{{\bm m}_\nu} ~,
\label{A.6.a}
\end{equation}
where the multiplicative factor
\begin{equation}
K \ = \ \frac{ 1 }{ \Gamma\left(n+1-\frac{D}{2}+\frac{|{\bm m}_1|+\ldots+|{\bm m}_n|}{2}\right)} ~.
\end{equation}

In the second step we invoke a resolutions of unity $\int_{\mathbb{R}^D} d{\bm y}_\nu \proj{{\bm y}_\nu} = \mathbb{I}$, which
brings $Q$ to the form
\begin{equation} \label{Q2}
Q \ = \ K
\bra{{\bm y}_{n+1}}
\frac{1}{1 + \frac{\hat{\bm q}^2}{2}} \hat{\bm y}^{{\bm m}_n} \frac{1}{1 + \frac{\hat{\bm q}^2}{2}} \hat{\bm y}^{{\bm m}_{n-1}}
\dots
\frac{1}{1 + \frac{\hat{\bm q}^2}{2}} \hat{\bm y}^{{\bm m}_1} \frac{1}{1 + \frac{\hat{\bm q}^2}{2}}
\ket{{\bm y}_0} ~.
\end{equation}
With the use of the algebraic identity
\begin{equation}
[\hat{y}_j , F(\hat{\bm q})] \ = \ i \left. \frac{\partial F({\bm q})}{\partial q_j} \right|_{{\bm q}=\hat{\bm q}}
~,
\end{equation}
and the fact that  $\hat{y}_j \ket{{\bm y}_0} = 0$, ($j= 1, \ldots, D$)  we can bring (\ref{Q2}) to the form (recall the definition ${\bm y}_0 = {\bm y}_{n+1} = {\bm 0}$)
\begin{equation}
Q \ = \ K \bra{{\bm y}_{n+1}} G(\hat{\bm q}) \ket{{\bm y}_0}
\ = \ K \int_{\mathbb{R}^D} \frac{d{\bm q}}{(2\pi)^D}\ \! G({\bm q})
~,
\label{A.10.a}
\end{equation}
with $G({\bm q})$ defined as
\begin{equation}
G({\bm q}) \ = \
\left( \frac{i^{|{\bm m}_n|}}{1+\frac{{\bm q}^2}{2}} \frac{\partial^{|{\bm m}_n|}}{\partial {\bm q}^{{\bm m}_n}} \right)
\ldots
\left( \frac{i^{|{\bm m}_1|}}{1+\frac{{\bm q}^2}{2}} \frac{\partial^{|{\bm m}_1|}}{\partial {\bm q}^{{\bm m}_1}} \right)
\frac{1}{1+\frac{{\bm q}^2}{2}}
~.
\label{A.11a}
\end{equation}

Note that we could arrive at the same conclusion by employing in (\ref{A.6.a}) the spectral expansion of the
position operator (or better, its power) in both position and momentum representations, i.e.
\begin{eqnarray}
\hat{\bm y}^{{\bm m}} \ = \ \int_{\mathbb{R}^D} d{\bm y}_\nu |{\bm y}_\nu\rangle {\bm y}_\nu^{{\bm m}} \langle
{\bm y}_\nu | \ = \ \int_{\mathbb{R}^D} d{\bm q}_\nu |{\bm q}_\nu\rangle i^{|{\bm m}|}{\frac{\partial^{|{\bm m}|}}{\partial
{\bm q}_\nu^{{\bm m}}}} \langle {\bm q}_\nu |\, .
\end{eqnarray}
Ensuing lack of one $\delta$-function then causes the residual ${\bm q}$-integration in (\ref{A.10.a}).

\section{Structure of $Q({\bm m}_1,\ldots,{\bm m}_n)$}\label{App.2}

In this appendix we show that the number of terms  involved in evaluating  ${Q}({\bm m}_1, \ldots, {\bm m}_n)$
via (\ref{Q11})  grows  as (\ref{B.8b}).
%
%
We start by observing that the function $G({\bm q})$ in (\ref{A.11a}) can be written as a sum
\begin{equation} \label{Gsum}
G({\bm q}) \ = \ \sum_{{\bm r},s} a_{{\bm r},s} \frac{{\bm q}^{\bm r}}{\left( 1+\frac{{\bm q}^2}{2} \right)^{\! s}}\, ,
\end{equation}
with combinatorial factors $a_{{\bm r},s}$ whose explicit form is not not relevant for the arguments to follow. The components of multi-index ${\bm r}$ satisfy $0 \leq r^j \leq m_1^j+\ldots+m_n^j$ for all $j=1,\ldots,D$ since each differentiation $\partial/\partial q_j$ can produce at most one power of $q_j$.

The summation index $s$ in (\ref{Gsum}) is not independent variable but it is fully specified once ${\bm r}$ is known. To see this, consider an elementary differentiation step
\begin{equation} \label{ElemDif}
\frac{\partial}{\partial q^j} \frac{{\bm q}^{\bm r}}{\left( 1+\frac{{\bm q}^2}{2} \right)^{\! s}}
\ = \ \frac{r^j {\bm q}^{{\bm r}-{\bm e}_j}}{\left( 1+\frac{{\bm q}^2}{2} \right)^{\! s}}
- \frac{s {\bm q}^{{\bm r}+{\bm e}_j}}{\left( 1+\frac{{\bm q}^2}{2} \right)^{\! s+1}}\,  ,
\end{equation}
where $e_j^i =\delta_{i j}$, and define $\Delta$ to be the difference between the degree of the polynomial in the denominator and the numerator. The derivative shifts $\Delta$ from $2s-|{\bm r}|$ to $2s-|{\bm r}|+1$, and this is common to both terms on the right hand side.  Hence, the nonzero terms in sum (\ref{Gsum}) must satisfy the condition $2s-|{\bm r}|=|{\bm m}_1|+\ldots+|{\bm m}_n|+2n+2$. This is also evident on the dimensional ground.

We also note that due to (\ref{ElemDif}) $r^j$ in (\ref{Gsum}) has, for all $j$, the same even parity as the total degree of differentiation $m_1^j+\ldots+m_n^j$ because otherwise the integral in (\ref{A.10.a}) would vanish. Altogether, we see that there are only
\begin{equation}
\sum_{{\bm r},s} 1 \ = \ \prod_{j=1}^D \ \! \sum_{r^j =0}^{m_1^j+\ldots+m_n^j} 1  \ = \ \prod_{j=1}^D \left[ (m_1^j+\ldots+m_n^j)/2+1 \right],
\end{equation}
non-trivially contributing terms in (\ref{Gsum}).

Let us close this appendix by proving the inequality (\ref{47.a}). To this end we observe that one can
write
\begin{eqnarray}
(2\mathfrak{m} - 1)!! \ &=& \ \frac{2\mathfrak{m}!}{2^\mathfrak{m} \mathfrak{m}! }
 = \  \frac{1}{\sqrt{\pi}} \ \! \Gamma \!\left[ 1/2 + \textstyle{\sum_{j=1}^D} (m_1^j+\ldots+m_n^j) \right] \prod_{j=1}^D 2^{m_1^j+\ldots+m_n^j}\nonumber \\[2mm]
 &\geq& \ \prod_{j=1}^D 2^{m_1^j+\ldots+m_n^j} \ \geq \ \prod_{j=1}^D\left[1 + (m_1^j+\ldots+m_n^j)/2\right] .
\end{eqnarray}
On the first line we have used the duplication formula~\cite{Gradshteyn}: $\Gamma(z) \Gamma(z+1/2) = \sqrt{\pi} \ \! \Gamma(2z) 2^{1-2z}$. On the second line the use was made of the inequality $\Gamma(1/2 + z) \geq \sqrt{\pi}$ (valid for $z\geq 2$) and the convexity inequality $2^z-1 \geq z \log 2 > z/2$ (valid for $z\geq 0$).


\begin{thebibliography}{25}


\bibitem{W-K} E.~Wigner, Phys. Rev.~{\bf40}, 749 (1932).

\bibitem{W-K2} J.G.~Kirkwood, Phys. Rev.~{\bf 44}, 31
(1933).

\bibitem{Simon:79} B.~Simon,  \emph{Functional Integration and Quantum Physics}, (Academic, New York, 1979).

\bibitem{Haba:99} Z.~Haba, \emph{Feynman Integral and Random Dynamics in Quantum Physics;
A Probabilistic Approach to Quantum Dynamics},
(Kluwer, London, 1999).

\bibitem{Kleinert} H.~Kleinert, \emph{Path Integrals in Quantum Mechanics, Statistics, Polymer Physics, and Financial Markets}, 5-th edition, (World Scientific, London, 2009).

\bibitem{zinn-justin} J.~Zinn-Justin, {\em Path Integrals in Quantum Mechanics}, (Oxford University Press, Oxford, 2005).

\bibitem{fischer:92} W.~Fischer, H.~Leschke and P.~M\"{u}ller, J.~Phys.~A:~Math.~Gen.~{\bf 25}, 3835 (1992) 3835.

\bibitem{Ceperley:2013} E.W.~Brown, B.K.~Clark, J.L.~DuBois and D.M.~Ceperley,
Phys.~Rev.~Lett.~{\bf 110}, 146405 (2013).

\bibitem{Krauth:2006} W.~Krauth,  \emph{Statistical Mechanics: Algorithms and Computations}, (Oxford University Press, Oxford, 2006)

\bibitem{Hioe:75} F.T.~Hioe and E.W. Montroll, J.~Math.~Phys. {\bf 16}, 1945 (19750.

\bibitem{Hioe:78} F.T.~Hioe, D.~Machlillen and E.W.~Montroll, Phys.~Rep.~{\bf 43}, 305 (1978).

\bibitem{Banerjee:78} K.~Banerjee, P.~Bhatnager, V. Choudry and S.S.~Kanwal, Proc.~Roy.~Soc.~(London)~A{\bf 360}, 575 (1978).

\bibitem{Witschel:80} W.~Witschel, Chem.~Phys.~{\bf 56}, 265 (1980).

\bibitem{Schwarz} M.~Schwartz,Jr., J.~Stat.~Phys.~{\bf 15}, 255 (1976).

\bibitem{Miller:71} W.H.~Miller, J.~Chem.~Phys.~{\bf 55}, 3146 (1971).

\bibitem{Jorish:75} V.S.~Jorish, V.Yu.~Zitserman, Chem.~Phys.~Lett.~{\bf 34},  378 (1975).

\bibitem{Korsch:79} H.J.~Korsch, J.~Phys~A:~Math.~Gen.~{\bf 12}, 1521 (1979).

\bibitem{Brantut} P.~Brantut, et al., Science~{\bf{337}},  1069 (2012).

\bibitem{Grenier} Ch.~Grenier, C.~Kollath and A. Georges, [ArXiv preprint 1209.3942].

\bibitem{Chabrier:06} G.~Chabrier,  D.~Saumon and A.~Potekhin, J.~Phys.~A{\bf~39}, 4411 (2006).

\bibitem{Martin} A.~Alastuey and Ph.A.~Martin, Phys.~Rev.~A{\bf 40}, 6485 (1989); B.~Jancovici, Mol~Phys.~{\bf 32}, 1177 (1976); M.M.~Gombert and D.~L\'{e}ger, Phys.~Rev.~E{\bf 57}, 3962 (1998).

\bibitem{Boyd:68} M.E.~Boyd, S.Y.~Larsen and J.E.~Kilpatrick, J.~Chem.~Phys.~{\bf 50}, 4034 (1969).

\bibitem{Cao:94} J.~Cao and G.A.~Voth, J.~Chem.~Phys.~{\bf 101}, 6168 (1994).

\bibitem{Militzer:01} B.~Militzer and D.M.~Ceperley, Phys.~Rev.~E{\bf~63}, 066404 (2001).

\bibitem{Predescu:03}  C.~Predescu, D.~Sabo and J.D.~Doll, J.~Chem.~Phys.~{\bf~119}, 4641 (2003).

\bibitem{Feynman-Kleinert} R.~Feynman and H.~Kleinert, Phys.~Rev.~A{\bf~34}, 5080 (1986).

\bibitem{Giachetti} R.~Giachetti and V.~Tognetti, Phys.~Rev.~Lett.~{\bf~55}, 912 (1985).

\bibitem{Angel} S.~Paulin, A.~Alastuey and T.~Dauxois, J.~Stat.~Phys. {\bf 128}, 1391 (2007).

\bibitem{schubert} C.~Schubert, Physics Report~{\bf 355}, 73 (2001).

\bibitem{Fliegner-I} D.~Fliegner, M.G.~Schmidt and C.~Schubert,  Z.~Phys.~C{\bf~64}, 11 (1994).

\bibitem{Fliegner-II} D.~Fliegner, P.~Haberl, M.G.~Schmidt and C.~Schubert, Ann.~Phys.~(N.Y.)~{\bf 264}, 51 (1998).

\bibitem{Belkov:93} A.A.~Belkov, D.~Ebert, A.V.~Lanyov and A.~Schaale, Int.~J.~Mod.~Phys.~A{\bf~8}, 1313 (1993).

\bibitem{Carso:90} L.~Carson, Phys.~Rev.~D{\bf~42}, 2853 (1990).

\bibitem{Kikkawa:76} K.~Kikkawa, Prog.~Theor.~Phys.~{\bf 56}, 947 (1976).

\bibitem{MacKenzie:84} R.~MacKenzie, F.~Wilczek and A.~Zee, Phys.~Rev.~Lett.~{\bf 53}, 2203 (1984).

\bibitem{Fraser:85} C.M.~Fraser, Z.~Phys.~C{\bf~28}, 101 (1985).

\bibitem{Gibson:1984} W.G.~Gibson, J.~Phys.~A:~Math.~Gen.~{\bf 17}, 1891 (1984).

\bibitem{FH} R.P.~Feynman and A.R.~Hibbs, \textit{Quantum Mechanics and Path
Integrals}, (McGraw-Hill, New York, 1965)

\bibitem{simons} B.~Simon, \textit{Functional Integration and Quantum Physics}, (Academic Press, New York, 1979).

\bibitem{Onofri:78} E.~Onofri, Am.~J.~Phys.~{\bf 46}, 379 (1978).

\bibitem{Zuk:85} J.A.~Zuk, J.~Phys.~A{\bf~18}, 1795 (1985).

\bibitem{Vladimirov} see, e.g., V.S.~Vladimirov, \textit{Equations of Mathematical Physics}, (Marcel Dekker, Inc, New York, 1971).

\bibitem{Jan78} B. Jancovici, Physica A, {\bf 91} , 152 (1978).

\bibitem{Kiha:24} T.~Kihara, Y.~Midzuno and T.~Shizume, J. Phys. Soc. Jpn.~{\bf 10}, 249 (1955).

\bibitem{landau} L.D.~Landau, and E.~Lifschitz, \textit{Statistical Physics}, (Pergamon, Elmsford, 1980).

\bibitem{Gradshteyn} I.S.~Gradshteyn and I.M.~Ryzhik, {\it Table of Integrals, Series, and Products}, 7-th edition, (Elsevier, New York, 2007).

\bibitem{fn1}
See supplementary material at [...] for {\scshape Wolfram Mathematica} notebook {\ttfamily expansion1D.nb} that implements the formulas (\ref{Q3}) and (\ref{1D}), and allows to insert a specific (smooth) potential $V(x)$. The expansion up to order $\beta^{18}$ is available in the file {\ttfamily expansion1D18orders.m}, and is imported automatically into the {\ttfamily .nb} notebook.

\bibitem{Bern-Kosower} Z.~Bern and D.A.~Kosower, Phys.~Rev.~Lett. {\bf~66},  1669 (1991); Nucl.~Phys.~B{\bf~379}, 451 (1992).

\bibitem{Strassler}  M.J.~Strassler, Nucl. Phys. B{\bf~385}, 145 (1992).

\bibitem{Schubert.3} D.~Fliegner, P.~Haberl, M.G.~Schmidt, C.~Schubert, Ann.~Phys.~(N.Y.) {\bf 264}, 51 (1998).

\bibitem{Schubert.2} D.~Fliegner, M.G.~Schmidt and C.~Schubert, Z.~Phys.~C{\bf~64}, 111 (1994).

\bibitem{Ala79} A. Alastuey and B. Jancovici, Physica A: Stat. Mech.~{\bf 97}, 349 (1979).

\bibitem{Wang:87} L.~Wang, Plasma~Phys.~Control.~Fusion~{\bf29}, 395 (1987).

\bibitem{Caro:96} J.~Caro, E.~Ruiz Arriola and L.L.~Salcedo, J.~Phys.~G{\bf~22}, 981 (1996).

\bibitem{Osborn}  T.A.~Osborn and F.H.~Molzahn, Phys.~Rev.~A{\bf 34}, 1696 (1986).

\bibitem{Durand} M.~Durand, P.~Schuck and X.~Vi\~{n}as, Phys.~Rev.~A{\bf 36}, 1824 (1987).

\bibitem{kleinert_b} H.~Kleinert, Phys.~Lett.~A{\bf~173}, 332 (1993).

\bibitem{Bender} C.M.~Bender, K.A.~Milton, M.~Moshe, S.S.~Pinsky and L.M.~Simmons Jr., Phys.~Rev.~Lett.~{\bf 58}, 2615 (1987).

\bibitem{Stevenson} P.M.~Stevenson, Phys.~Rev.~D{\bf 23}, 2916 (1981).

\bibitem{BJV} M.~Blasone, P.~Jizba and G.~Vitiello, {\it Quantum Field Theory and its Macroscopic Manifestations} (Imperial College Press, London, 2011).






\end{thebibliography}
\end{document}